\documentclass[twocolumn,showpacs,superscriptaddress]{revtex4}
\usepackage{graphicx}
\usepackage{natbib}
\usepackage{latexsym}
\begin{document}

\newcommand{\dum}{$\left.\right.$}


\title
{
Interferometry in dense nonlinear media and interaction-induced loss
of contrast in microfabricated atom interferometers
}
\author{Maxim Olshanii}
\affiliation{Permanent Address: Department of Physics \& Astronomy,
University of Southern California, Los Angeles, CA 90089}
\affiliation{ITAMP, Harvard, Cambridge, Massachusetts 02138}
\author{Vanja Dunjko}
\affiliation{Permanent Address: Department of Physics \& Astronomy,
University of Southern California, Los Angeles, CA 90089}
\affiliation{ITAMP, Harvard, Cambridge, Massachusetts 02138}
\date{\today }

\begin{abstract}
In this paper we update the existing schemes for computation of
atom-interferometric signal in single-atom interferometers to
interferometry with dense Bose-condensed atomic samples. Using the
theory developed we explain the fringe contrast degradation
observed, for longer duration of interferometric cycle, in the
Michelson interferometer on a chip recently realized at JILA
(Ying-Ju Wang, Dana Z. Anderson, Victor M. Bright, Eric A.
Cornell, Quentin Diot, Tetsuo Kishimoto, Mara Prentiss, R. A.
Saravanan, Stephen R. Segal, Saijun Wu,  Phys. Rev. Lett. {\bf
94}, 090405 (2005)). We further suggest several recipes for
suppression of the interaction-related contrast degradation.

\end{abstract}

\maketitle

{\it Introduction}.-- Atom interferometers
\cite{mlynek,pritchard,kasevich,berman_review} offer an
unprecedented precision in inertial measurements. Supplemented
with a highly coherent input source provided by Bose-condensed
atoms
\cite{bill,hagley,kasevich_disks,ertmer_experiment,immanuel_lattice,
dave_double_well},
atom interferometers may potentially supersede the conventional
laser-based devices. In a recent experiment \cite{dana} a
miniature Michelson-type interferometer was realized on an atom chip
\cite{hinds,danlop,jorg,zimmermann,reichel,mara},
thus further approaching practical implementations of the device.
Generally, miniaturization of atomic devices leads to an increased
role of interatomic interactions, due to higher densities and
density gradients. Indeed, a strong suppression of contrast was
observed in \cite{dana} for longer durations of the
interferometric cycle. The goal of our paper is to explain this
effect and suggest recipes for suppressing the interaction-related
fringe degradation.

The role of interatomic interactions in interferometric processes
has been studied by several authors
\cite{martin,marvin,haldane,zozulya_instability}, with the main
emphasis on the potential loss of first-order coherence. In our
paper we focus on a different effect: distortion of the
interferometric path due to the mean-field pressure.

\begin{figure}
%
\includegraphics[scale=.4]{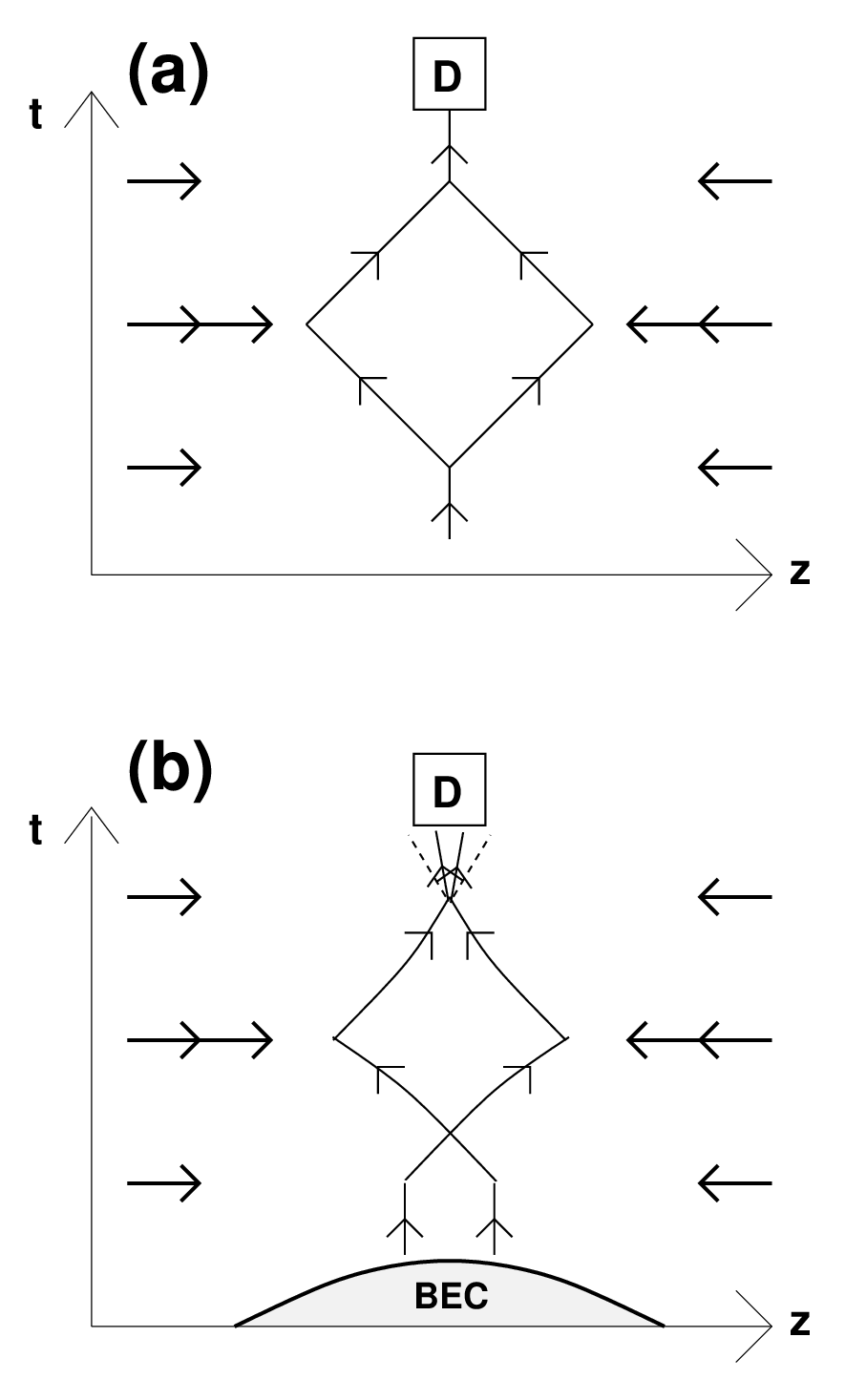}
\caption { \label{f:scheme} Interferometric loop of the Michelson
interferometer with single atoms (a) and with Bose condensates
(b). }
\end{figure}

{\it Interferometric scheme}.-- In this paper we consider the
Michelson interferometric scheme (see Fig.\ \ref{f:scheme}(a)).
Atoms are supposed to be confined transversally by a 
monomode atom guide, with no transverse excitations allowed.
The initial state of atoms is a perfectly coherent state
\begin{eqnarray}
\psi(z,\,t=0-) = \chi(z)
\quad,
\label{psi(0-)}
\end{eqnarray}
normalized for convenience to the total number of atoms 
$N$: $\int_{-\infty}^{+\infty} dz\, |\chi(z)|^2 = N$.
We further assume that at every stage of the process
the wave function can be approximately decomposed
into a sum of three spatial harmonics,
\begin{eqnarray}
\psi =
\frac{1}{\sqrt{2}}\sum_{n=-1,\,0,\,+1} \Phi_{n}(z,\,t) \,
e^{inQz}
\quad,
\end{eqnarray}
where
$\Phi_{n}(z,\,t)$ are slow functions of coordinate. (Note that
even though higher harmonics can be generated during the
interferometric process, we show below that they are (a) small
under typical conditions, and (b) if necessary can be taken into
account {\it a posteriori}.) Splitting, reflection, and
recombining pulses perform the following instant transformations
of the vector $(\Phi_{-1},\,\Phi_{0},\,\Phi_{+1})$:
\begin{eqnarray}
&&
\hat{A}_{\text{split.}} = \hat{A}_{\text{rec.}} = 
\frac{1}{2}
\left(
\begin{array}{ccc}
1        & \sqrt{2} & -1
\\
\sqrt{2} & 0        & \sqrt{2}
\\
-1       & \sqrt{2} & 1 
\end{array}
\right) 
\nonumber
\\
&&
\hat{A}_{\text{refl.}} = \left(
\begin{array}{ccc}
0 & 0 & 1
\\
0 & 0 & 0
\\
1 & 0 & 0
\end{array}
\right)
\label{elements}
.
\end{eqnarray}
Such ideal interferometric elements were proposed in
\cite{beamsplitter} and successfully experimentally realized in
\cite{dana}. The splitting, reflection, and recombination pulses
are applied in succession, separated by an equal time interval
$T$.
Immediately after the recombination
pulse the population of atoms in the central peak is detected;
this constitutes the interferometric signal.

Between the splitting and
recombination pulses the wave function can be
approximately decomposed
as
\begin{eqnarray}
&&\psi(z,\,t) \approx
\nonumber
\\
&&
\quad
\frac{1}{\sqrt{2}}
\left\{
e^{i\phi_{+}(z,\,t)} e^{i(m\bar{v}_{+}(t)z-\bar{\epsilon}_{+}(t)t)/\hbar}
\chi(z-\bar{z}_{+}(t))
\right.
\nonumber
\\
&&
\quad
\left.
+
e^{i\phi_{-}(z,\,t)} e^{i(m\bar{v}_{-}(t)z-\bar{\epsilon}_{-}(t)t)/\hbar}
\chi(z-\bar{z}_{-}(t))
\right\}
,
\label{psi_basic}
\end{eqnarray}
where the phases $\phi_{\pm}$ will be shown to be approximately real. Here
\begin{eqnarray}
\bar{z}_{\pm}(t) =
\left\{
\begin{array}{lll}
\pm V_{Q} t                 & \mbox{for}  & 0<t<T
\\
\pm V_{Q} T \mp V_{Q} (t-T) & \mbox{for}  & T<t<2T
\end{array}
\right.
\label{classical_coordinate}
\end{eqnarray}
are the classical trajectories, originating at $z=0$,
corresponding to the right (+) and left (-) arms of the
interferometer;
\begin{eqnarray}
\bar{v}_{\pm}(t) = \dot{\bar{z}}_{\pm}(t) =
\left\{
\begin{array}{lll}
\pm V_{Q} & \mbox{for}  & 0<t<T
\\
\mp V_{Q} & \mbox{for}  & T<t<2T
\end{array}
\right.
\label{classical_velocity}
\end{eqnarray}
are the corresponding classical velocities; finally,
$\bar{\epsilon}_{+}(t) = \bar{\epsilon}_{-}(t) = E_{Q} \equiv
mV_{Q}^2/2$ are the corresponding kinetic energies. The velocity
$V_{Q}$ is given by $V_{Q} \equiv \hbar Q /m$, where $m$ is the
atomic mass.

The resulting differential phase shift
\begin{eqnarray}
%
\Delta\phi(z) \equiv \phi_{+}(z,\,2T) -\phi_{-}(z,\,2T)
\label{differential_phase_shift}
\end{eqnarray}
consists of two parts: $\Delta\phi(z) = \Delta\phi_{\text{signal}}
+ \Delta\phi_{\text{distortion}}(z)$. The first, spatially
independent part is the useful signal, related to the effect the
interferometer measures. The second part is the result of the
distortion caused by unaccounted-for external fields and, in our
case, mean-field interactions. The distortion phase shift leads to
two effects. The first is a correction to the signal phase shift.
This effect can in principle be accounted for, if the nature of
the distortion is known. The second effect is a degradation of
contrast in the interferometric signal. This degradation can not
be eliminated easily if $\Delta\phi_{\text{distortion}}(z)$
changes substantially over the length of the atomic cloud. If both
effects are taken into account, then the interferometric signal
\begin{eqnarray*}
&&S(\Delta\phi_{\text{signal}}) \equiv N^{-1} \times
\\
&&\quad \int_{-\infty}^{+\infty} dz\,
\cos^2((\Delta\phi_{\text{signal}} +
\Delta\phi_{\text{distortion}}(z))/2) |\chi(z)|^2 
\end{eqnarray*}
can be shown to be
\begin{eqnarray*}
S(\Delta\phi_{\text{signal}}) = \frac{1}{2} + \frac{M}{2}
\cos(\Delta\phi_{\text{signal}}-\delta) \quad,
\end{eqnarray*}
where the fringe contrast $M$ and fringe shift $\delta$ are given by
\begin{eqnarray}
&&M = \sqrt{A^2+B^2}
\label{fringe_contrast}
\\
&& \delta = \arg(A,\,B)
\label{fringe_shift}
\quad,
\end{eqnarray}
and
\begin{eqnarray*}
&&A = N^{-1} \int_{-\infty}^{+\infty} dz\,
\cos(\Delta\phi_{\text{distortion}}(z)) |\chi(z)|^2 
\\
&&B= N^{-1} \int_{-\infty}^{+\infty} dz\,
\sin(\Delta\phi_{\text{distortion}}(z)) |\chi(z)|^2 
\quad.
\end{eqnarray*}
The goal of this paper is to calculate the fringe contrast
degradation caused by mean-field interactions, compare the results
with the experimentally observed values, and suggest methods for
eliminating this degradation.

{\it Analysis of the mean-field effects}.--
We describe the evolution of the
atomic cloud using the nonlinear
Schr\"{o}dinger equation
\begin{eqnarray}
i\hbar\frac{\partial}{\partial t}\psi
+
\frac{\hbar^2}{2m}\frac{\partial^2}{\partial z^2} \psi
=
U(z,\,t) \psi
+
g_{\rm 1D}|\psi|^2 \psi
\quad,
\end{eqnarray}
where $U(z,\,t)$ is an external field comprising the object of the
measurement and any other auxiliary fields present; $
g_{\rm 1D}=-\hbar^{2}/\mu\,a_{\rm 1D} $ is the one-dimensional coupling
constant; $ a_{\rm 1D} = (-a^{2}_{\perp}/2 a)\,[1-{\mathcal
C}(a/a_{\bot})] $ is the one-dimensional scattering length; $\mu =
m/2$ is the reduced mass; $ a_{\perp}=\sqrt{\hbar / \mu
\,\omega_{\perp}} $ is the size of the transverse ground state of
the guide; $ {\mathcal C} = 1.4603\dots $ (see \cite{Olshanii});
lastly, $a$ is the three-dimensional $s$-wave scattering length.
We further decompose the wave function into a quasi-Fourier series
\begin{eqnarray}
\psi =
\frac{1}{\sqrt{2}}\sum_{n=\pm 1,\, \pm 3, \ldots} \psi_{n}
\quad,
\end{eqnarray}
where each term obeys
\begin{eqnarray}
&&\left(i\hbar\frac{\partial}{\partial t}
+
\frac{\hbar^2}{2m}\frac{\partial^2}{\partial z^2}\right) \psi_{n}
=
\label{quasi-Fourier}
\\
&&\quad
U(z,\,t) \psi_{n}
+
\frac{g_{\rm 1D}}{2} 
\sum_{n_2,\,n_3} \psi^{\star}_{n_2+n_3-n}\psi_{n_2}\psi_{n_3}
\quad.
\nonumber
\end{eqnarray}
The initial condition for the system (\ref{quasi-Fourier}) is
given by the result of applying the beamsplitter (\ref{elements})
to the initial condition (\ref{psi(0-)}). We get
\begin{eqnarray*}
&&\psi_{\pm 1}(z,\, t=0+) = \chi(z) e^{\pm Qz}
\\
&&\psi_{n \neq \pm 1}(z,\, t=0+) = 0
\quad.
\end{eqnarray*}

Let us assume that the $\psi_{\pm 1}$ components of the wave
function remain dominant, as they are initially, and then verify
this assumption for self-consistency. Here and throughout the
paper we will suppose that the beamsplitter recoil energy $E_{Q}$
is much greater than the mean-field energy $g_{\rm 1D}|\chi|^2$:
\begin{eqnarray}
\varepsilon_{{\rm nl}} \sim \frac{g_{\rm 1D}|\chi|^2}{E_{Q}}
\ll 1
\quad.
\label{eps_nl}
\end{eqnarray}
Under these assumptions the strongest higher harmonics generated
are
\begin{eqnarray}
\psi_{\pm 3}
=
-\frac{1}{16} 
\frac{g_{\rm 1D}\psi^{\star}_{\mp 1}\psi_{\pm 1}}{E_{Q}} \psi_{\pm 1}
\quad,
\label{third_harmonics}
\end{eqnarray}
and it is indeed small 
provided the parameter (\ref{eps_nl}) is
small. Another assumption used to derive (\ref{third_harmonics})
was that the beamsplitter recoil energy $E_{Q}$ dominates another
energy scale associated with the deviation of the momentum
distribution of the dominant harmonics $\psi_{\pm 1}$ from that of
strict $\delta$-peaks around $p=\pm\hbar Q$. As we will see from
the following, in a realistic experiment this energy scale
satisfies an even stricter requirement of being dominated by the
``time-of-flight'' energy $\hbar/T$.

In general, according to our findings the dynamics of higher
harmonics is simply reduced to an adiabatic following of the
dominant $\psi_{\pm 1}$ ones. This observation greatly simplifies
further analysis. In short, one may safely exclude the couplings
to higher harmonics from the equations of evolution for $\psi_{\pm
1}$, obtain results, and add the correction
(\ref{third_harmonics}) {\it a posteriori}.

The equations for the dominant harmonics $\psi_{\pm 1}$
now become
\begin{eqnarray}
\left(i\hbar\frac{\partial}{\partial t}
+
\frac{\hbar^2}{2m}\frac{\partial^2}{\partial z^2}\right) \psi_{\pm 1}
=
U_{\pm}(z,\,t) \psi_{\pm 1}
\quad,
\label{psi_pm_equation}
\end{eqnarray}
where the effective potentials for the
right (+) and left (-) arms of the interferometer
read
\begin{eqnarray}
U_{\pm}(z,\,t) = U(z,\,t) + \frac{g_{\rm 1D}}{2} |\psi_{\pm 1}|^2
                          + g_{\rm 1D} |\psi_{\mp 1}|^2
\quad.
\label{potential}
\end{eqnarray}
%
\begin{figure}
%
\includegraphics[scale=.4]{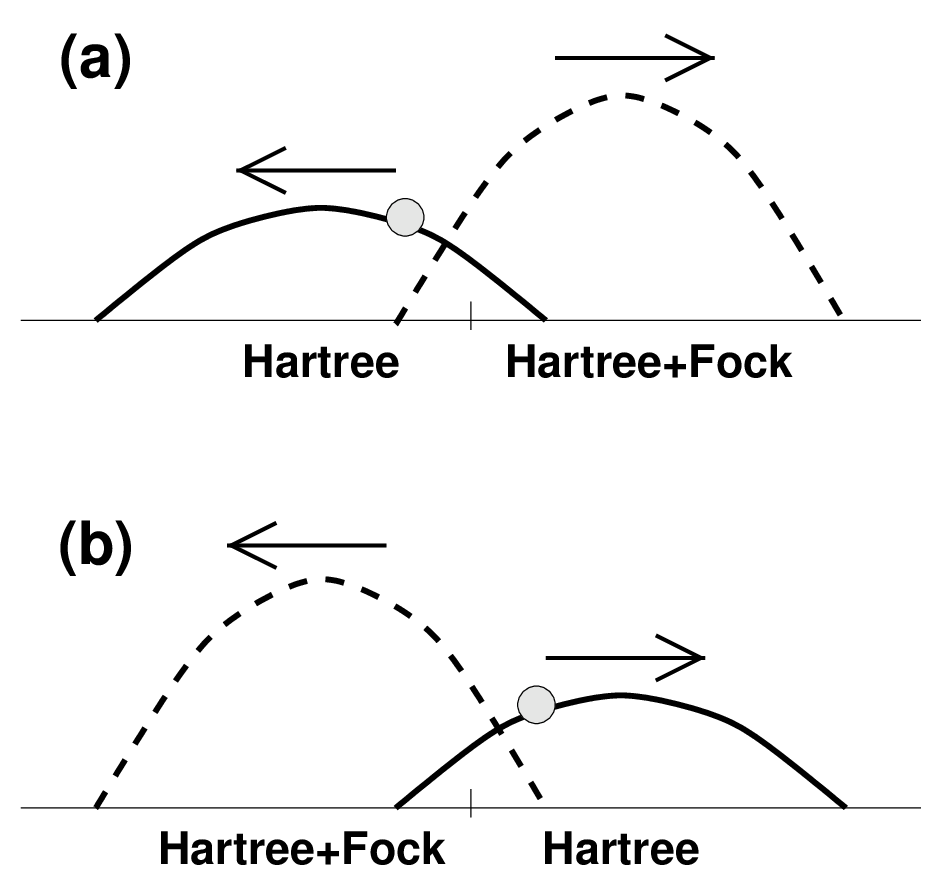}
\caption { \label{f:mf_pressure} An artistic view of the
mean-field effects in a Michelson interferometer. Notice that the
mean-field potential is different for the right arm (a) and the
left arm (b). }
\end{figure}
The first part of the mean-field potential is the contribution of
the part of the cloud to which the given interferometer arm
belongs. The second, twice-as-strong part is the influence of the
opposite arm. The factor of two difference between the two
contributions (illustrated in Fig.\ \ref{f:mf_pressure}) can be
traced to the difference between the Hartree-type interaction
among atoms in the condensate, and the Hartree-Fock-type
interaction between atoms in the condensate and those out of the
condensate.

In what follows we study the effect of the
mean-field interaction (\ref{potential}) on the
differential phase shift (\ref{differential_phase_shift})
and fringe contrast (\ref{fringe_contrast}).

{\it Differential phase shift.}-- From now on
we will be using the representation (\ref{psi_basic}) for the wave
function, where the phase factors $\phi_{\pm}(z,\,t)$ (generally
complex) are still unknown. The equation (\ref{psi_pm_equation})
under the representation (\ref{psi_basic}) leads to the following
equations for the phase factors
\begin{eqnarray}
\frac{\partial}{\partial \tau_{\pm}} \phi_{\pm} = -
U_{\pm}(z,\,t)/\hbar + D_{\pm}(z,\,t) \quad. \label{phi_equation}
\end{eqnarray}
Here
\begin{eqnarray}
\frac{\partial}{\partial \tau_{\pm}} \equiv
\frac{\partial}{\partial t} +
\bar{v}_{\pm}(t)\frac{\partial}{\partial z} \quad;
\end{eqnarray}
the classical velocity $\bar{v}_{\pm}$ is given in
(\ref{classical_velocity}); the potentials (\ref{potential}) are
\begin{eqnarray}
&&U_{\pm}(z,\,t) = U(z,\,t)
\\
&&\quad + \frac{g}{2} |\chi(z-\bar{z}_{\pm}(t))|^2 + g
|\chi(z-\bar{z}_{\mp}(t))|^2 \quad;
\nonumber \label{potential_2}
\end{eqnarray}
the classical coordinate $\bar{z}_{\mp}(t)$ is represented by
(\ref{classical_coordinate}). Finally, the correction term $D$ is
\begin{eqnarray}
D_{\pm} = D_{J,\,\pm} + D_{{\rm dyn.},\,\pm} + D_{{\rm dyn.-init.},\,\pm} + D_{{\rm init.},\,\pm}
\quad,
\label{correction_1}
\end{eqnarray}
where
\begin{eqnarray}
&&D_{J,\,\pm} = i \frac{\hbar^2}{2m} \frac{\partial^2}{\partial
z^2} \phi_{\pm}\,,
\\
&&D_{{\rm dyn.},\,\pm} = -\frac{\hbar^2}{2m}
\left(\frac{\partial}{\partial z}\phi_{\pm}\right)^2\,,
\\
&&D_{{\rm dyn.-init.},\,\pm}\,,
\\
&&\quad = i\frac{\hbar}{m} \left(\frac{\partial}{\partial
z}\phi_{\pm}\right) \left(\frac{\partial}{\partial
z}\ln(\chi(z-\bar{z}_{\pm}(t)))\right)\,,\;\text{and}
\nonumber
\\
&&
D_{{\rm init.},\,\pm} = \frac{\hbar^2}{2m} \frac{\frac{\partial^2}{\partial z^2}\chi(z-\bar{z}_{\pm}(t))}{\chi(z-\bar{z}_{\pm}(t))}
\quad.
\label{correction_2}
\end{eqnarray}

Now we decompose the phase shift
\begin{eqnarray}
\phi_{\pm} = \bar{\phi}_{\pm} + \delta\phi_{\pm}
\end{eqnarray}
into a sum of the principal part obeying
\begin{eqnarray}
\frac{\partial}{\partial \tau_{\pm}} \bar{\phi}_{\pm} = - U(z,\,t)_{\pm}/\hbar
\quad,
\label{phi_bar_equation}
\end{eqnarray}
and the correction originating from the $D$ term in
(\ref{phi_equation}). We assume that the correction
$\delta\phi_{\pm}$ is small,
\begin{eqnarray}
\delta\phi_{\pm} \ll 1
\quad,
\label{phi_correction}
\end{eqnarray}
which we will justify later.

Finally the
differential phase shift reads
\begin{eqnarray}
&&\Delta\phi(z) = -\hbar^{-1}
\int_{0}^{2T} dt' \times
\label{differential_phase_shift_2}
\\
&&\quad\quad
\left\{
U_{+}(z-\bar{z}_{+}(2T)+\bar{z}_{+}(t'),\,t')
\right.
\nonumber
\\
&&\quad\quad\quad
\left.
-
U_{-}(z-\bar{z}_{-}(2T)+\bar{z}_{-}(t'),\,t')
\right\}
\nonumber
\end{eqnarray}
In what follows we will assume that 
the initial atomic wave function has the
Thomas-Fermi profile
\begin{eqnarray}
\chi(z) = \frac{m\omega_{\rm BEC}^2}{2} (R^2-z^2) \theta(R-|z|)
\label{Thomas-Fermi}
\end{eqnarray}
Having in mind certain applications we will also add
an external harmonic oscillator potential
\begin{eqnarray}
U(z,\,t) = \frac{m\omega_{0}^2}{2} z^2
\end{eqnarray}
of an arbitrary frequency $\omega_{0}$. We will further classify both 
as distortion effects and set the useful signal 
to zero. To the order of approximations we made, the 
caused by distortion fringe shift and fringe contrast degradation 
analyzed below are independent of the useful signal, and thus the latter 
can indeed be omitted. The distortion differential phase shift 
now becomes equal to the total one: $\Delta\phi_{\text{distortion}}(z) = 
\Delta\phi(z)$. 
 
{\it Interaction-induced loss of contrast: small interferometers}.--
First consider the case of small interferometers, where 
the interferometric arms are substantially overlapped 
during the whole interferometric cycle, $V_{Q}T \ll R$.
In this case we can easily
compute the combined mean-field and harmonic-oscillator
contribution to the differential phase shift $\Delta\phi$ (see
(\ref{differential_phase_shift_2})), as well as the resulting fringe
contrast $M$ (see \ref{fringe_contrast}). They read
\begin{eqnarray}
\Delta\phi_{\text{distortion}}(z) = 2Kz \quad \label{result_phi}
\end{eqnarray}
and
\begin{eqnarray}
M = |f(2\tilde{K}R)| \quad.
 \label{result_contrast}
\end{eqnarray}
Here
\begin{eqnarray}
\tilde{K} = m \tilde{\omega}^2 T^{2} V_{Q}  / \hbar
\label{K}
\end{eqnarray}
has the meaning of the half of the differential momentum acquired by atoms 
in the mean field and in the harmonic oscillator field 
(see Fig.\ \ref{f:scheme}(b)),
\begin{eqnarray}
\tilde{\omega}^2 = 2\omega_{\rm BEC}^2 - \omega_{0}^2
\label{omega_tilde}
\quad,
\end{eqnarray}
and
\begin{eqnarray}
f(\xi) = 3 (\sin(\xi) -\xi\cos(\xi))/\xi^3
\label{function_f}
\quad.
\end{eqnarray}
Here and below $\omega_{\rm BEC}$ is the frequency of the harmonic
trap for which the state (\ref{Thomas-Fermi}) would be the ground
state, and $R$ is the Thomas-Fermi radius. Note that for the
configuration considered the fringes are suppressed, but 
either not
shifted at all or shifted by $\pi$ 
(see (\ref{fringe_shift})). Notice also that for a particular case 
of $\omega_{0} = \sqrt{2}\omega_{\rm BEC}$ the differential phase 
shift vanishes and thus the fringe contrast is strictly 100\%.
We will discuss the implications of this phenomenon below.

The expression (\ref{result_contrast}) for the fringe contrast
in small interferometers
is the {\it first principal result} of this paper.

\begin{figure}
%
\includegraphics[scale=.32]{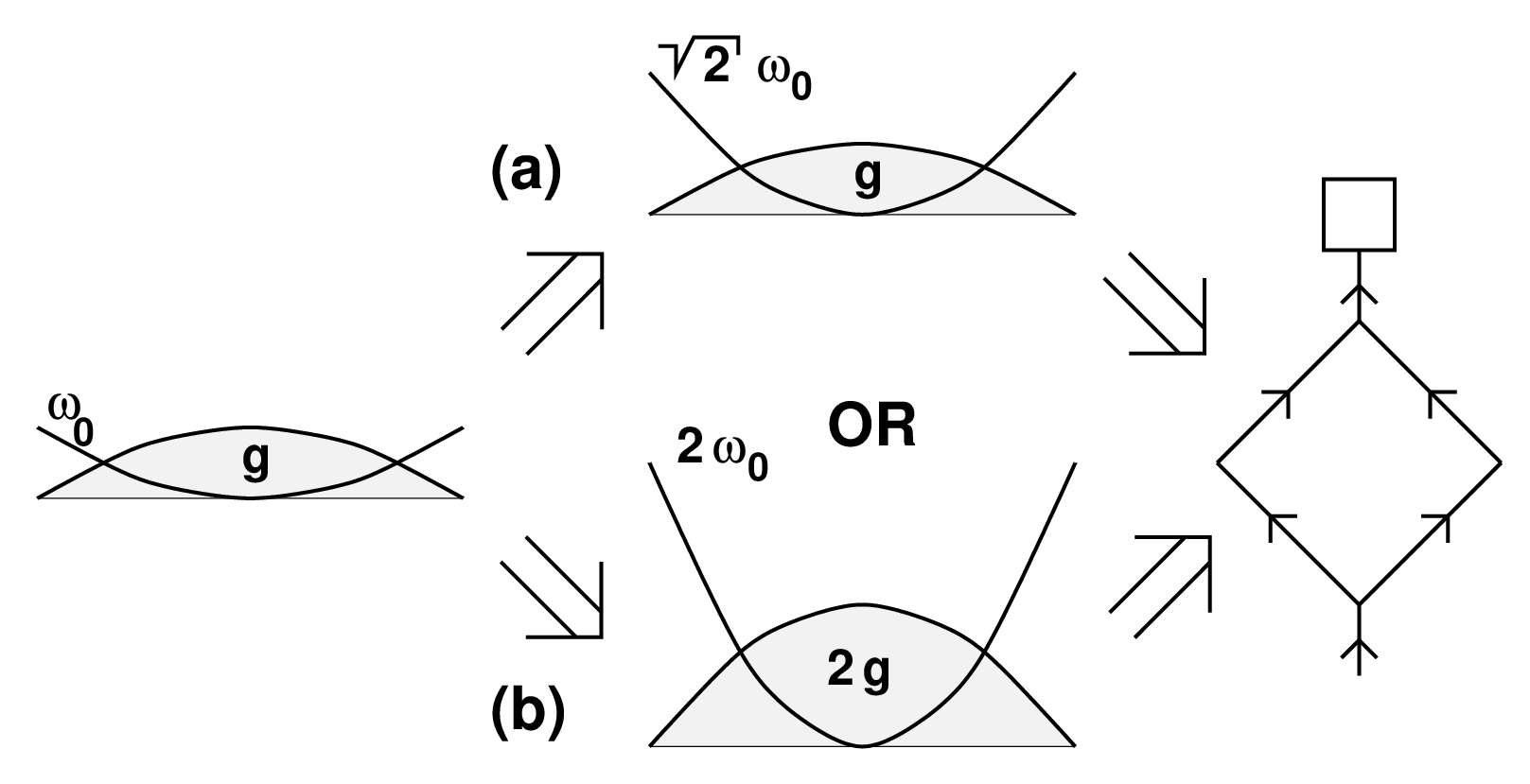}
\caption { \label{f:remedies} Two schemes for the preparation of
the initial wave packet, designed to eliminate the 
interatomic-interaction-induced
degradation of the fringe contrast in small interferometers. 
(a) Situation when the
longitudinal confinement can be controlled independently from the
transverse one. (b) Situation when the longitudinal and transverse
confinements are linearly linked one to another. }
\end{figure}

{\it Remedies for the contrast degradation: small interferometers}.--
As one can see from
(\ref{result_phi}), the distortion of the differential phase shift
in small interferometers
disappears completely if the frequency of the external harmonic
trap is chosen to be $\sqrt{2}$ higher than the frequency of the
mean-field potential:
\begin{eqnarray}
\omega_{0} = \sqrt{2}\, \omega_{\rm BEC} \quad\Rightarrow\quad
\Delta\phi_{\text{distortion}} = 0 \quad \label{remedy_condition}.
\end{eqnarray}
This can be realized in two ways:

(a) The first scheme is applicable if the longitudinal frequency
$\omega$ can be controlled independently of the transverse
frequency $\omega_{\perp}$. In this case, one should start by
preparing the condensate in the ground state of the longitudinal
trap. Then, just before the splitting pulse, one should increase
the longitudinal frequency by a factor of $\sqrt{2}$ in a short
ramp (Fig.\ \ref{f:remedies}(a)).

(b) The second scheme assumes that both longitudinal and
transverse frequencies are controlled by the same source, and thus
the ratio between them is always the same. Then the change in the
longitudinal frequency will affect the nonlinear coupling constant
$g$ (to which the square of the condensate frequency $\omega_{\rm
BEC}^2$ is linearly proportional) due to  the simultaneous change
in the transverse frequency. Note that $g$ is linearly
proportional to $\omega_{\perp}$. In this case one satisfies the
condition (\ref{remedy_condition}) by increasing, prior to the
splitting pulse, both the longitudinal and the transverse
frequencies by a factor of 2 (Fig.\ \ref{f:remedies}(b)).

Note that both schemes are completely {\it insensitive} to
possible fluctuations in the number
of particles in the condensate.

{\it Interaction-induced loss of contrast: large interferometers}.--
Consider now the the case of large interferometers where
for some period of time during the  cycle the arms are 
totally spatially separated: $V_{Q}T \ge R$. The differential 
phase shift in this case reads  
\begin{eqnarray}
\Delta\phi_{\text{distortion}}(z) = 
2K_{0}z 
-z^3/l^3
\quad, 
\label{result_phi_large}
\end{eqnarray}
where
\begin{eqnarray}
&&K_{0} = - m \omega_{0}^2 T^{2} V_{Q}  / \hbar
\\
&&l = (m \omega_{\rm BEC}^2 /3\hbar V_{Q})^{\frac{1}{3}}
\quad.
\label{K_large}
\end{eqnarray}
Notice that unlike in the small interferometer case, for no 
choice of parameters the differential phase shift can be completely 
eliminated, and that the contribution from the harmonic 
potential grows with the time duration of the cycle while 
the one from the mean-field is stationary. This makes us to believe 
that for large interferometers the most promising implementation 
will be the free-space one with no longitudinal confinement 
present.    

\begin{figure}
%
\includegraphics[scale=.7]{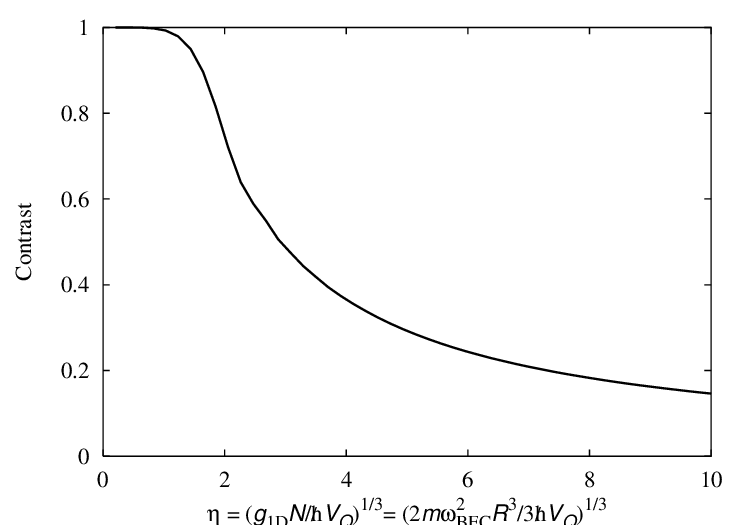}
\caption 
{ 
\label{f:contrast_ratio_large} Fringe contrast ratio vs.\
the universal parameter $\eta$ for large interferometers.
}
\end{figure}

In the case of $\omega_{0} = 0$ the fringe contrast assumes 
the following compact expression:
\begin{eqnarray}
M = {\cal F}(\eta) 
\quad,
\label{result_contrast_large}
\end{eqnarray}
(see Fig.\ \ref{f:contrast_ratio_large}) where
\begin{eqnarray}
\eta &=& (2m\omega_{\rm BEC}^2 R^3/3\hbar V_{Q})^{\frac{1}{3}}
\nonumber
\\
     &=& (g_{\rm 1D} N/\hbar V_{Q})^{\frac{1}{3}} 
\quad,
\label{eta}
\end{eqnarray}
is the parameter governing the fringe contrast,
\begin{eqnarray}
&&{\cal F}(\eta) = \frac{3}{2} {_1F_2}(1/6;\, 1/2,\, 7/6;\, -\eta^6/16) 
\nonumber
\\
&&\qquad -\sin(\eta^3/2)/\eta^3
\quad,
\label{result_function_F}
\end{eqnarray}
and $_nF_m(a_1,\,\ldots,\,a_n;\, b_1,\,\ldots,\,b_m;\, \xi)$ 
is the generalized hypergeometric function.

Notice that the parameter $\eta$ depends on neither 
the shape of the atomic cloud nor on the interferometric cycle duration.
For a given set of atomic and waveguide parameters the 
large contrast requirement $\eta \ll 1$  defines a 
{\it universal limit} for the number of atoms:
\begin{eqnarray}
N \ll \hbar V_{Q}/g_{\rm 1D}
\quad,
\label{universal_limit}
\end{eqnarray}

The expression (\ref{result_contrast_large}) 
is the {\it second principal result} of our paper.

{\it Limits of the validity of our computational scheme}.-- In
order for our conclusions be valid the correction
(\ref{phi_correction}) originating from the neglected terms in the
kinetic energy must be small. We have performed a thorough
investigation aimed at understanding the physical meaning of the
neglected corrections and estimating their value. The results are
as follows.

The correction $\delta\phi_{\pm}$ can be decomposed into
a sum of four terms
\begin{eqnarray}
&&\delta\phi_{\pm} = \delta\phi_{J,\,\pm} + \delta\phi_{{\rm dyn.},\,\pm} +
\nonumber
\\
&&\qquad
\delta\phi_{{\rm dyn.-init.},\,\pm} + \delta\phi_{{\rm init.},\,\pm}
\label{phi_correction_2}
\end{eqnarray}
with the following interpretation:
\begin{eqnarray}
&&
e^{\delta\phi_{J,\,\pm}} \approx |J_{U_{\pm}}|^{-\frac{1}{2}}
\\
&&
e^{\delta\phi_{{\rm dyn.},\,\pm} + \delta\phi_{{\rm dyn.-init.},\,\pm} + \delta\phi_{{\rm init.},\,\pm}}
\\
&&\quad
\approx
\langle
\chi(z-\bar{z}_{\pm}(t))
|
e^{-i\int_{0}^{t} dt \, (p_{U_{\pm}}+\hat{p})/2m\hbar}
|
\chi(z-\bar{z}_{\pm}(t))
\rangle
.
\nonumber
\end{eqnarray}
The first correction is related to the expansion factor (Jacobian)
of the bunch of trajectories of classical particles moving in the
field $U_{\pm}$. The second, third, and forth corrections
originate from the neglected kinetic energy, both the initial
kinetic energy (coming from the momentum distribution of $\chi$)
and that acquired in the field $U_{\pm}$. 
The above
corrections (together with the nonlinear correction
(\ref{eps_nl})) lead to the following requirements for the validity
of the approximations used:
\begin{eqnarray}
&&\varepsilon_{J} = (\tilde{\omega}T)^2 \ll 1
\\
&&\varepsilon_{{\rm dyn.}} = m\tilde{\omega}^4 \tilde{z}^2  T^3 /
\hbar \ll 1
\\
&&\varepsilon_{{\rm init.}} = \hbar T / m R^2 \ll 1
\\
&&\varepsilon_{{\rm nl}} = (\tilde{\omega} R/ V_{Q})^2 \ll 1
\quad,
\label{eps_lin_1}
\end{eqnarray}
where $\tilde{z}$ is the typical atomic coordinate, 
given by $\tilde{z} \sim R$ ($\tilde{z} \sim V_{Q}T$) for small (large)
interferometers. 
For a typical set  of parameters of the JILA experiment 
with 
$\tilde{\omega} = 2\pi \times 3.2 \mbox{Hz}$ and $T=10^{-3}\,\mbox{s}$
(corresponding to 
the small interferometer case), the values of these
parameters are indeed small, validating our approximation:
$\varepsilon_{J} = \varepsilon_{{\rm dyn.-init.}} = 4.1 \times
10^{-4}$, $\varepsilon_{{\rm dyn.}} = 4.7 \times 10^{-4}$,
$\varepsilon_{{\rm init.}} = 3.6 \times 10^{-4}$, and
$\varepsilon_{{\rm nl}} = 3.6 \times 10^{-3}$.

\begin{figure}
%
\includegraphics[scale=.7]{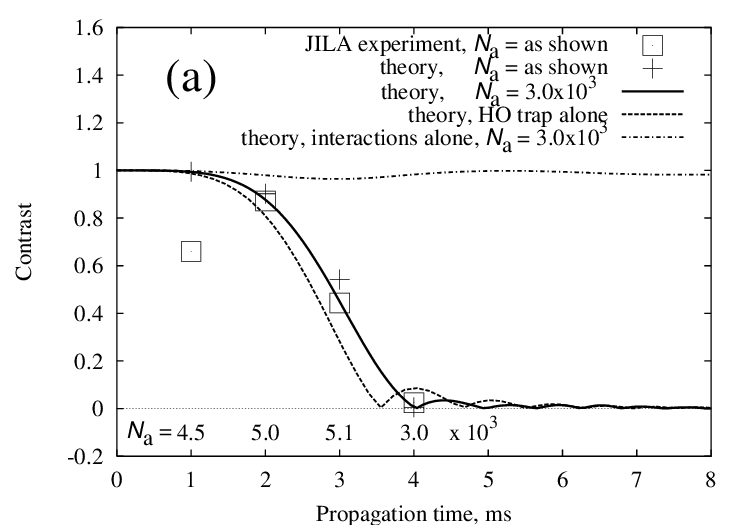}
\includegraphics[scale=.7]{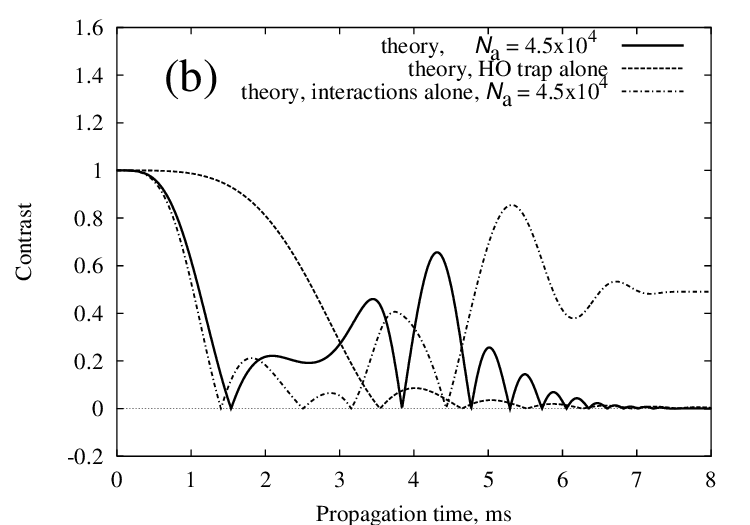}
\caption { \label{f:contrast_ratio_general} Fringe contrast ratio 
vs.\
duration of the interferometric cycle corresponding to 
the parameters of the JILA experiment with
Michelson interferometer on a chip, with magnetic gradient as the
phase element. (a) Curves correspond to $N=3.\times 10^{3}$ atoms,
the as same 4 ms point in the experiment. The theoretical predictions 
for the actual experimental numbers of atoms at every run are also 
shown. (b) The same as (a), but for $N=4.5\times 10^{4}$ atoms,
where a reduction of the fringe contrast to 50\% is expected.}
\end{figure}

{\it Comparison with the JILA experiment}.--
The parameters of the JILA's experiment on 
Michelson interferometer on a magnetic
chip \cite{dana} lie in the range intermediate between 
the small and large interferometer regimes and requires no 
assumption on the distance between the arms $V_Q$ {\it vis a vis} 
the cloud size $R$. The equation for the fringe contrast in this 
case must be integrated numerically. 
In \cite{dana} a time- and
space-localized pulse of magnetic field was used as the phase
signal. A stationary harmonic potential of frequency 
$\omega_{0} = 2\pi \times 5\, \mbox{Hz}$ was present in each 
realization. The contrast was traced as a function of the duration of
the interferometric cycle $2T$, as depicted in Fig.\
\ref{f:contrast_ratio_general}. Other parameters read 
$\omega_{\perp} = 2\pi \times 100\, \mbox{Hz}$,
$Q=4\pi/\lambda$, where
$\lambda = 780\, \mbox{nm}$ is the wavelength of light used to
produce the interferometric elements, $R=45\,\mu\text{m}$, and
$a=100.4\, a_{\text{B}}$. 
In the experiment the number of atoms varied from one
value of the cycle duration to another; these numbers are shown in
Fig.\ \ref{f:contrast_ratio_general}(a). The value of $\omega_{\rm BEC}$ was
extracted from $\omega_{\rm BEC}^2 =
(4/9)g_{\rm 1D}N/mR^3$. The results of the
comparison are shown at the Fig.\ \ref{f:contrast_ratio_general}(a).

One can see from the Fig.\ \ref{f:contrast_ratio_general} 
that for the parameters chosen the role of interatomic interactions 
in contrast degradation is relatively small and the main 
source of the effect is the stationary harmonic trap. This is entirely 
unexpected since the strength of the interactions was very 
close to the strength of the trap, typically $\omega_{\rm BEC}^2 =
.25 \div .4 \omega_{0}^2$, and moreover in the small interferometer 
regime 
the strength interactions become multiplied by a factor 
of 2 of Fock origin (see (\ref{omega_tilde}). 
One can show further that the relatively weak role of 
interactions in the JILA experiment is not related to any 
small parameter, but is solely an interplay of numerical 
prefactors. To illustrate this point we show at the 
Fig.\ \ref{f:contrast_ratio_general}(b) a theoretical prediction for 
$N = 4.5 \times 10^{4}$ atoms exhibiting a 
dominant role of the interatomic interactions. 

{\it Summary and outlook.}-- 

(1) We have developed a simple computational scheme 
that allows to include the effect of interatomic interactions 
in calculation of fringe shift and fringe degradation in
waveguide based atom interferometers.

(2) In two cases we have found 
simple analytic expression for fringe contrast. These cases are:
small interferometers where the spatial separation between the 
arms is much smaller than the atomic cloud size, $V_{Q}T \ll R$;
and large interferometers where at at least one instance 
the arms are fully separated, $V_{Q}T \ge R$.

(3) In the case of small interferometers the analytic expression for 
the contrast degradation allowed us to suggest a simple recipe 
canceling the destructive effect of interactions {\it completely}.
 
(4) In the case of large interferometers the effect of interactions 
can not, to our knowledge, be canceled entirely. Furthermore,
the interatomic interactions set an {\it universal limit} on the 
number of atoms involved in the interferometric process:
$N \ll \hbar V_{Q}/g_{\rm 1D}$. Notice that this bound depends on the 
characteristics of the atom and the waveguide only (where 
the ``beam-splitter velocity'' $V_{Q}$ is supposed to be linked 
to the atomic transition frequency), while neither 
the timing of cycle nor the size of the atomic cloud enter.    

(5) Using the method developed we have analyzed the results of the 
recent JILA experiment on Michelson interferometer on 
atom chip. In spite of comparable strength of interatomic interactions
and stationary longitudinal trap present in the experiment,
our results indicate a relatively weak role of interactions  
in fringe contrast degradation. This finding is can not be traced to 
any small parameter, but is a mere interplay of numerical 
prefactors. A moderate ten-fold increase (to $N = 4.5 \times 10^{4}$) 
in the number 
of atoms will reduce, for large interferometer case,  
the fringe visibility to 50\%, even without 
an additional longitudinal trap present, reflecting the 
universal limit outlined above.

(6) Michelson interferometer is a closed-loop white-light scheme
by design; it is supposed to produce clear fringes even for input
sources with a short coherence length. As one can see from the
Fig.\ \ref{f:mf_pressure}(b), the mean-field pressure leads to two
distinct effects. The first effect is the change in the relative
momentum of the interferometer arms; this is what we addressed in
the present work. The second effect is the distortion of the
interferometric path, as a result of which the path becomes open,
and thus the interferometer becomes sensitive to the longitudinal
coherence. For zero-temperature condensates this does not lead to
any loss of contrast. At finite temperature the degradation due to
the broken interferometer loop becomes relevant, and we are going
to study this effect in the nearest future.

\begin{acknowledgments}
We are grateful to Ying-Ju Wang and Dana Z. Anderson for providing
us with the recent experimental data and for enlightening
discussions on the subject. This work was supported by a grant
from Office of Naval Research {\it N00014-03-1-0427}, and through
the National Science Foundation grant for the Institute for
Theoretical Atomic and Molecular Physics at Harvard University and
Smithsonian Astrophysical Observatory.
\end{acknowledgments}


\begin{thebibliography}{}
%
\bibitem{berman_review} {\it Atom Interferometry},
edited by P. R. Berman (Academic, New York, 1997).
%
%
\bibitem{mlynek} O. Carnal and J. Mlynek, Phys. Rev. Lett. {\bf 66}, 2689 (1991).
%
\bibitem{pritchard} D. W. Keith, C. R. Ekstrom, Q. A. Turchette, and D. E. Pritchard,
Phys. Rev. Lett. {\bf 66}, 2693 (1991).
%
\bibitem{kasevich} M. Kasevich and S. Chu, Phys. Rev. Lett. {\bf 67}, 181 (1991).
%
%
\bibitem{bill} J. E. Simsarian, J. Denschlag, M. Edwards, C. W. Clark, L. Deng,
E. W. Hagley, K. Helmerson, S. L. Rolston, and W. D. Phillips,
Phys. Rev. Lett. {\bf 85}, 2040 (2000).
%
\bibitem{hagley}Y. Torii, Y. Suzuki, M. Kozuma, T. Sugiura,
T. Kuga, L. Deng, and E. W. Hagley, Phys. Rev. {\bf A61}, 041602 (2000).
%
\bibitem{kasevich_disks} B. P. Anderson, M. A. Kasevich, Science {\bf 282} 1686 (1998).
%
\bibitem{ertmer_experiment} D. Hellweg, L. Cacciapuoti, M. Kottke, T. Schulte,
K. Sengstock, W. Ertmer, J. J. Arlt, Phys. Rev. Lett. {\bf 91}, 010406 (2003).
%
\bibitem{immanuel_lattice} Artur Widera, Olaf Mandel, Markus Greiner, Susanne Kreim, Theodor W. Hänsch, and Immanuel Bloch,
Phys. Rev. Lett. {\bf 92}, 160406 (2004).
%
\bibitem{dave_double_well} Y. Shin, M. Saba, T. A. Pasquini, W. Ketterle, D. E. Pritchard,
A. E. Leanhardt, Phys. Rev. Lett. {\bf 92}, 050405 (2004).
%
%
\bibitem{dana} Ying-Ju Wang, Dana Z. Anderson, Victor M. Bright, 
Eric A. Cornell,
Quentin Diot, Tetsuo Kishimoto, Mara Prentiss,
R. A. Saravanan, Stephen R. Segal, Saijun Wu,  Phys. Rev. Lett. {\bf 94}, 
090405 (2005).
%
%
\bibitem{hinds}  P. K. Rekdal, S. Scheel, P. L. Knight, E. A. Hinds,
Phys. Rev. {\bf A 70}, 013811 (2004).
%
\bibitem{danlop}  C. J. Vale, B. Upcroft, M. J. Davis, N. R. Heckenberg, 
H. Rubinsztein-Dunlop, J.Phys. {\bf B 37} 2959 (2004).
%
\bibitem{jorg} S. Schneider, A. Kasper, Ch. vom Hagen, M. Bartenstein, 
B. Engeser, T. Schumm, I. Bar-Joseph, R. Folman, L. Feenstra, 
and J. Schmiedmayer, Phys. Rev. {\bf A 67}, 023612 (2003). 
%
\bibitem{zimmermann} H. Ott, J. Fortagh, G. Schlotterbeck, A. Grossmann, 
and C. Zimmermann,  
Phys. Rev. Lett. {\bf 87}, 230401 (2001). 
%
\bibitem{reichel} W. Hänsel, P. Hommelhoff, T. W. Hänsch, 
and J. Reichel, Nature (London) {\bf 413}, 498 (2001). 
%
\bibitem{mara}
M. Vengalattore, W. Rooijakkers, and M. Prentiss, 
Phys. Rev. {\bf A 66}, 053403 (2002).
%
%
\bibitem{martin} A. Rohrl, M. Naraschewski, A. Schenzle, H. Wallis,
Phys. Rev. Lett. {\bf 78}, 4143 (1997).
%
\bibitem{marvin} M. D. Girardeau, K. K. Das, E. M. Wright,
Phys. Rev. {\bf A66}, 023604 (2002).
%
\bibitem{haldane} S. Chen and R. Egger, Phys. Rev. {\bf A68}, 063605 (2003).
%
\bibitem{zozulya_instability} J. A. Stickney and A. A. Zozulya, Phys. Rev. {\bf A 66},
053601 (2002).
%
%
\bibitem{beamsplitter}
Saijun Wu, Yingju Wang, Quentin Diot, Mara Prentiss, e-print physics/0408011.
%
\bibitem{Olshanii} M. Olshanii, Phys. Rev. Lett. {\bf 81}, 938 (1998).
\end{thebibliography}
\end{document}